\documentclass[%
 superscriptaddress,
preprint,
showpacs,preprintnumbers,
nofootinbib,
 amsmath,amssymb,
 aps,
 prd,
]{revtex4-1}

\pdfoutput=1

\usepackage{amssymb}
\usepackage{mathrsfs}
\usepackage{dsfont}
\usepackage{graphicx,subfigure}
\usepackage{enumitem}

\usepackage[colorlinks,
            linkcolor=black,
            anchorcolor=black,
            citecolor=black
            ]{hyperref}

\usepackage{xcolor}
\usepackage{multirow}

\newlength{\dinwidth}
\newlength{\dinmargin}
\newcommand{\GeV}{{\,\rm GeV}}
\newcommand{\diff}{{\rm d}}

\begin{document}
\preprint{NCTS-PH/1801}

\title{Search for light sterile neutrinos from $W^\pm$ decays at the LHC}
\date{}

\author{Claudio O. Dib}
\email[E-mail: ]{claudio.dib@usm.cl}
\affiliation{Department of Physics and CCTVal, Universidad T\' ecnica Federico Santa Mar\'\i a,
Valpara\'\i so  2340000, Chile}

\author{C.S. Kim}
\email[E-mail: ]{cskim@yonsei.ac.kr}
\affiliation{Department of Physics and IPAP, Yonsei University, Seoul 120-749, Korea}
\affiliation{Institute of High Energy Physics, Dongshin University, Naju 58245, Korea}
\author{Sebastian Tapia Araya}
\email[E-mail: ]{s.tapia@cern.ch}
\affiliation{Department of Physics, University of Illinois at Urbana-Champaign, Urbana, IL 61801, USA}

\date{\today}

\begin{abstract}
\noindent
We study the feasibility to observe sterile neutrinos with masses in the range
\hbox{5~GeV~$< m_N < 20$~GeV} at the LHC, using exclusive semileptonic modes involving pions, $W\to \ell N \to \pi \ell\ell, 2\pi \ell\ell$
and $3\pi \ell\ell$. We thus cover a mass window that is between what can be studied in meson factories and high energy colliders. We run simulations for these exclusive events, where pions should be distinguished from the background provided the neutrino decay exhibits a vertex displacement from its production point. In a previous work we have estimated the theoretical rates and here we analyze the observability of the processes at the LHC, given the fact that exclusive hadronic states may be difficult to identify.
 We study the sensitivity bounds for the observation and discovery of sterile neutrinos in the above mass range.
By the end of Run-3,  current bounds on heavy-to-light lepton mixings in the lower mass end ($\sim 5$ GeV) could be improved by about an order of magnitude to $|U_{\ell N}|^2\sim 5\times 10^{-6}$, and the High-Luminosity LHC could reach $|U_{\ell N}|^2 \lesssim 3\times 10^{-7}$ in the mass range below 11 GeV.
 Studying in addition equal sign and opposite sign dileptons, the Majorana or Dirac character of the sterile neutrino could be revealed.

\end{abstract}

\maketitle


\section{Introduction}

The discovery of neutrino oscillations ~\cite{Fukuda:1998mi,Ahmad:2002jz,Eguchi:2002dm} implies that at least two of the three neutrinos that participate in the weak interactions must be massive. Since neutrinos are massless in the Standard Model, these particles have become an important portal to physics beyond
the Standard Model~\cite{GonzalezGarcia:2007ib,Mohapatra:2006gs,Barger:2003qi,Strumia:2006db}. Now, the smallness of the observed neutrino masses is usually explained by introducing right-handed neutrinos, which must be sterile under electroweak interactions, inducing scenarios with a seesaw mechanism~\cite{Minkowski:1977sc,Yanagida:1979as,Ramond:1979py,GellMann:1980vs,Glashow:1979nm,Mohapatra:1979ia,Schechter:1981cv}. While originally the seesaw mechanism resorted to the presence of very heavy extra neutrinos, there are also models where the extra fields are not so heavy, leaving the open possibility that the extra sterile neutrinos could have masses in the broad range from eV to TeV~\cite{Deppisch:2015qwa}, and so experimental searches must also cover all those possibilities. Indeed there are plenty of scenarios in the literature to explain the light neutrino masses. These masses can be generated at tree level or  as loop contributions, and they almost invariably require the inclusion of extra fields. The so called type I, II and III include fermion singlets, scalar triplets and fermion triplets, respectively. Loop generated masses also include extra fields, typically scalars and/or fermions. Some of the scenarios may contain candidates for Dark Matter. In summary, in order to discriminate from the different scenarios that extend the Standard Model it will be important to know at least whether there exist extra neutral fermions, and in the affirmative case, to know their masses and whether they are Majorana or Dirac particles.

LHC searches for sterile neutrinos with mass above $100\GeV$~\cite{Chatrchyan:2012fla,Aad:2015xaa} are based on the inclusive processes
$pp \to W^\ast X$, $W^\ast\to \ell^\pm  \ell^\pm jj$~\cite{Keung:1983uu, delAguila:2007qnc, Atre:2009rg, Das:2017gke}. For $m_N$ below $M_W$, the jets are not energetic enough to pass the background reduction cuts, so purely leptonic modes $W^{(\ast)}\to \ell\ell \ell\nu $ could be preferred~\cite{Cvetic:2012hd, Dib:2015oka,Izaguirre:2015pga, Dib:2016wge,Dib:2017vux,Dib:2017iva,Dube:2017jgo}, even though they have the problem of missing energy and flavor number due to the undetectable final neutrino. However, as one goes to lower $m_N$ searches, again low $p_t$ leptons plus missing neutrinos affect the observability of these leptonic decays.
Now, for neutrino masses below $20$ GeV, there is an advantage: the neutrino may live long enough to leave an observable displacement from its production to its decay point~\cite{Helo:2013esa,Dib:2014iga,Gago:2015vma,Antusch:2017hhu,Cottin:2018kmq,Nemevsek:2018bbt, Cottin:2018hyf,Abada:2018sfh,Boiarska:2019jcw,Drewes:2019fou}, a feature that helps drastically reduce the backgrounds.
We then proposed to go back to using semileptonic modes for the searches in this mass range, but now with exclusive channels instead of jets~\cite{Dib:2018iyr}.
Again, for $m_N$  below 5 GeV, $B$ factories may be more appropriate to search for the sterile neutrino than high energy hadron colliders, due to the cleaner environment and the production of $N$ in $B$ or lighter meson decays~\cite{Cvetic:2010rw,Aaij:2012zr,Cvetic:2017vwl,Dib:2000wm,Kim:2017pra}.

For the sterile neutrino mass range of 5 GeV $<m_N< $ 20 GeV, in a  previous work~\cite{Dib:2018iyr} we proposed to use neither $\ell\ell jj$ nor trilepton events, but the exclusive semileptonic processes $W\to \mu N$, followed by $N \to \mu \pi$, $\mu \pi \pi$ and $\mu \pi\pi\pi$, which are modes with no missing energy. The decay channels $N\to e\pi, e\pi\pi, e\pi\pi\pi$ in the secondary process were not considered in order to avoid misidentification of electrons and pions. We concluded that the most promising modes should be $W^+\to \mu^+ N$ followed by a displaced decay
either $N\to \pi \mu^+$, $2\pi\mu^+$ or $3\pi\mu^+$  for a Majorana sterile neutrino, or $W^+\to \mu^+ N$ followed by $N\to \pi\mu^- $, $2\pi\mu^-$ or $3\pi\mu^-$ for a Dirac neutrino.
We studied those rates, including the comparison of different models for the pion form factors.

Now in this article we want to complement the previous work by studying the observability of these processes at the LHC. In general, the observability of these modes is not a trivial matter, since pions with relatively low $p_T$ need to be selected from backgrounds; neutral pions, which decay almost instantly into $\gamma\gamma$, are also difficult to identify; pions and electrons should be clearly distinguished in order to avoid fake signals. On the other hand, the vertex separation due to the sizable lifetime of the sterile neutrino with mass below 20 GeV~\cite{Helo:2013esa,Dib:2014iga,Gago:2015vma,Antusch:2017hhu,Cottin:2018kmq,Nemevsek:2018bbt, Cottin:2018hyf,Abada:2018sfh,Boiarska:2019jcw,Drewes:2019fou} can help reject considerably all backgrounds.

In Section II we review the processes in question in very brief form, as more details can be found in a previous work~\cite{Dib:2018iyr}. In Section III we present our current analysis and simulations, where we study the detectability of the processes at the LHC. In Section IV   we state our conclusions.

\section{Theoretical summary of the processes}\label{sec:decay}

  Here we give a short summary on our previous work on the decay $W \to \ell N$ and decays
  $N \to \ell n\pi$ with $n=1,2,3$. Detailed formulae with full theoretical discussions are shown in
  Ref.~\cite{Dib:2018iyr}.

\subsection{The decay $\boldsymbol{W \to \ell N}$:}

The leptonic sector in a generic SM extension includes one
or more extra neutral lepton singlets, $N$,
in addition to the three generations of left-handed SM $SU(2)_L$ lepton doublets.
  The neutral lepton singlets $N$ are
\emph{sterile}, in the sense that
they do not directly interact with other SM particles,
except through mixing with the active neutrinos.
At the LHC, sterile neutrinos with masses around $5\sim 20\GeV$ will be mainly produced from the decay of on-shell $W$ bosons. The decay rate $W^+\to \ell^+ N$ can be easily calculated; neglecting the charged  lepton mass, the branching ratio is: 
\begin{align}
  {\cal B}(W^+ \to \ell^+ N) \equiv \frac{\Gamma(W^+\to \ell^+ N)}{\Gamma_W} =  \frac{G_F}{\sqrt{2}} \frac{M_W^3}{12\pi \Gamma_W} |U_{\ell N}|^2 \left( 2+\frac{m_N^2}{M_W^2}\right)\left( 1- \frac{m_N^2}{M_W^2}\right)^2,
\end{align}
where $\Gamma_W \simeq 2.085$ GeV is the total decay width of the $W$ boson~\cite{Patrignani:2016xqp}.
From here, the neutrino $N$ can decay in several modes, depending on its mass. Here we are interested in the decays into pions, namely $N \to \pi^\mp \ell^\pm$, $N \to \pi^0 \pi^\mp \ell^\pm$  and $N \to \pi^\mp \pi^\mp \pi^\pm \ell^\pm$. Both charged modes will occur for a Majorana $N$, while for a Dirac $N$
only the $N$ decays into a negative charged lepton will be produced.

\subsection{The decay $\boldsymbol{N \to \pi^- \ell^+}$:}

The mode  $N \to \pi^\mp \ell^\pm$ is a charged current
process:
\begin{align}\label{eq:width:1pi}
\Gamma(N \to \pi^\mp \ell^\pm)=&\frac{G_F^2}{16\pi}f_\pi^2 \lvert V_{ud} \rvert^2 \lvert U_{\ell N} \rvert ^2\,  m_N^3 \lambda^{1/2} (1,m_\ell^2/m_N^2, m_{\pi^-}^2/m_N^2) \\
&\qquad\quad
\times
 \left[\left (1+ \frac{m_\ell^2}{m_N^2}-\frac{m_{\pi}^2}{m_N^2}\right) \left( 1+\frac{m_\ell^2}{m_N^2}\right) - 4\frac{m_\ell^2}{m_N^2}\right] ,
 \nonumber
\end{align}
where $m_{\pi}$ and $m_N$ denote the mass of the charged pion and sterile neutrino, respectively; $V_{ud}$ is the CKM matrix and $f_\pi$ is the pion decay constant; the function $\lambda (x,y,z)$ is defined as $\lambda(x,y,z)=x^2+y^2+z^2-2(xy+yz+zx)$.
The formation of a single pion in the final state is relatively suppressed with respect to multi pion modes, because it requires the two produced quarks to remain close together.
Indeed, the suppression relative to the open quark production is about $\sim 4 \pi^2 f_\pi^2/m_N^2$,
which is $\sim 0.6\%$ for $m_N= 10$ GeV \cite{Dib:2018iyr}.

\subsection{The decay $\boldsymbol{N \to \pi^0 \pi^- \ell^+}$:}\label{sec2b}

The decay into two pions, $N\to \pi^0 \pi^- \ell^+$, is similar to the $tau$ lepton decay $\tau^- \to \pi^0 \pi^- \nu_\tau$ in terms of their interaction lagrangian and Feynman diagram, except for the lepton flavor and charge.
However, one must be aware that the kinematic range for the form factor in the $N$ decays is extended to higher $q^2$, so an extrapolation of the $tau$ form factor will be required.
Considering the above, the differential decay rate for $N\to \pi^0 \pi^-\ell^+$ can be written as
\begin{align}\label{eq:width}
  \frac{\diff\Gamma(N \to \pi^0 \pi^- \ell^+)}{\diff s}=&\frac{\Gamma_N^0 \lvert V_{ud} \rvert^2 \lvert U_{\ell N} \rvert^2 }{2m_N^2} \frac{3s^3\beta_\ell \beta_\pi}{2m_N^6}F_-(s)^2 \\
&
\times
\left[ \beta_\ell^2 \left( \frac{(\Delta m_\pi^2)^2}{s^2} - \frac{\beta_\pi^2}{3}\right) + \left( \frac{(m_N^2 - m_\ell^2)^2}{s^2}-1\right) \left( \frac{(\Delta m_\pi^2)^2}{s^2} + \beta_\pi^2 \right) \right], \nonumber
\end{align}
where   $\Gamma_N^0 =G_F^2m_N^5/(192\pi^3)$, $s = (p_{\pi^0}+p_{\pi^+})^2$,
$\Delta m_\pi^2 \equiv m_{\pi^+}^2 - m_{\pi^0}^2$, $\beta_\ell =\lambda^{1/2}(1,m_\ell^2/s,m_N^2/s)$,
$\beta_\pi =\lambda^{1/2}(1,m_{\pi^+}^2/s,m_{\pi^0}^2/s)$, and
$F_-(s)$ is the hadronic form factor of the charged current, defined by
\begin{align}
  \langle \pi^-(p) \pi^0(p^\prime) \vert \bar d \gamma_\mu u \vert 0 \rangle &= \sqrt{2} F_-(s) (p-p^\prime)_\mu  .
\label{eq:FF:-}
\end{align}

The decay rate is then obtained after integrating over $s$,
within the limits $s_- = (m_{\pi^-} + m_{\pi^0})^2$ and $s_+ = (m_N - m_\ell)^2$.
This expression is analogous to  $\Gamma(\tau^- \to \pi^0 \pi^- \nu_\tau)$ \cite{Cirigliano:2001er,Cirigliano:2002pv}. The form factor $F_-(s)$ in the time-like region, i.e.\  $s>0$, is experimentally known from $\tau^- \to \pi^- \pi^0 \nu_\tau$~\cite{Fujikawa:2008ma} in the limited range $2m_\pi<\sqrt{s}<m_\tau$. The extrapolation to larger values of $s$ is done in our previous work \cite{Dib:2018iyr}, based on two alternatives: a vector dominance model~\cite{Lees:2012cj}
and on light front holographic QCD~\cite{Brodsky:2014yha}. They both give very similar
results~\cite{Dib:2018iyr}.

\subsection{The decay $\boldsymbol{N \to \pi^- \pi^- \pi^+ \ell^+}$:}

In much the same way as in the two-pion mode, 
the  differential decay rate of the general hadronic decay $N \to h_1 h_2 h_3 \ell^+$ can be written in terms of form factors with an expression identical to that of the tau decay $\tau^- \to h_1 h_2 h_3 \nu_\tau$ \cite{Dumm:2009kj}, again provided that the form factors are extrapolated to larger values of $q^2$.
Denoting the momentum and mass of the hadron $h_i$ ($i=1,2,3$) by $p_i$ and $m_i$ respectively, and defining the momentum of the hadronic part by
$q^\mu = (p_1+p_2+p_3)^\mu$,
the differential decay rate can be expressed as:
\begin{align}\label{eq:width:3h}
  \frac{\diff \Gamma(N \to h_1 h_2 h_3 \ell^+)}{\diff q^2} =& \frac{G_F^2|V_{ud}|^2|U_{\ell N}|^2}{128(2\pi)^5\ m_N^3}  \lambda^{1/2}(1, m_N^2/q^2 ,m_\ell^2/q^2)
    \bigg[ \left( \frac{(m_N^2-m_\ell^2)^2}{q^2} - m_N^2 - m_\ell^2 \right) \omega_{SA} (q^2)
  \nonumber \\
  &
    + \frac{1}{3}\left( \frac{(m_N^2 - m_\ell^2)^2}{q^2}+ m_N^2 + m_\ell^2  -2q^2\right)(\omega_A(q^2) + \omega_B(q^2)) \bigg] .
\end{align}

%

In this expression, the functions $\omega_A(q^2), \omega_B(q^2)$ and $\omega_{SA}(q^2)$ are given in Ref~\cite{Dib:2018iyr}.
The decay rate is obtained after integration over $q^2$ within the limits
$q_-^2=(m_1+m_2+m_3)^2$ and  $q_+^2 =(m_N-m_\ell)^2$.

\subsection{Theoretical results}

With the expressions described above we were able to estimate the exclusive semileptonic decay rates of $N$ into $\pi \ell$, $2\pi \ell$ and $3\pi\ell$, for a neutrino $N$ with mass in the range 5 to 20 GeV, produced at the LHC in the process $W \to \ell N$.
\begin{figure}[h]
  \centering
  \includegraphics[width=0.49\textwidth]{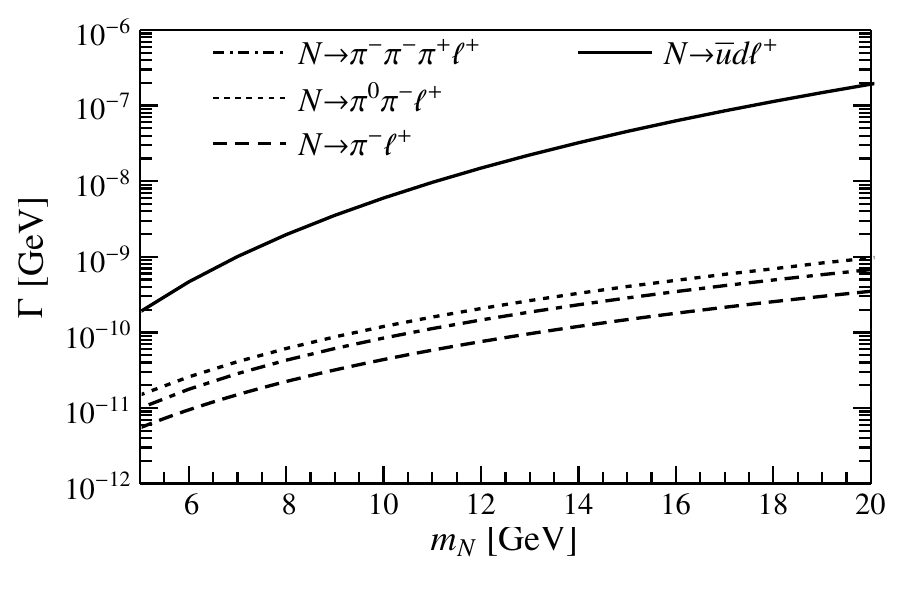}
  \includegraphics[width=0.49\textwidth]{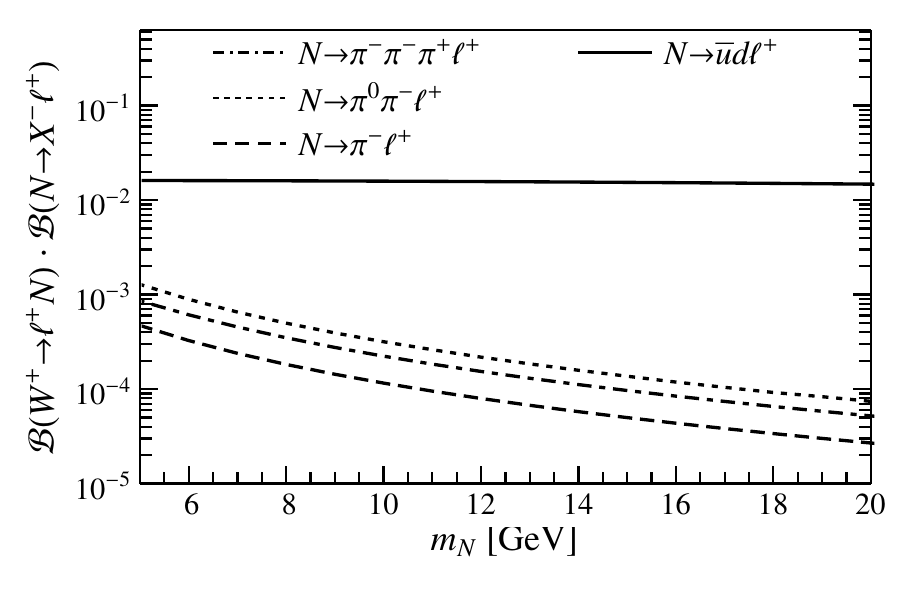}
  \caption{\baselineskip 3.0ex
 ``Canonical'' decay rates for $N \to \pi^- \ell^+$, $N \to \pi^0 \pi^- \ell^+$, $N \to \pi^- \pi^- \pi^+ \ell^+$ and the inclusive mode $N \to \bar u d \ell^+$ (left) and ``canonical'' branching ratios for the full processes $W^+\to \ell^+\ell^+ n \pi$ (right) as a function of the neutrino mass $m_N$, with all mixing factors $|U_{\ell N}|$ removed. To obtain the actual values, the ``canonical''  values must be multiplied by the factor
 $|U_{\ell N}|^2$ (left), or  $|U_{\ell N}|^4/\sum_{l_i} |U_{\ell_i N}|^2$ (right).
 }
 \label{fig:width:tot}
\end{figure}
In Fig.~\ref{fig:width:tot} (left) we reproduce what we called the ``canonical'' decay rates for the modes $N \to n \pi+ \ell^+$ and the inclusive estimate given by $N \to \bar u d \ell^+$  as a function of the neutrino mass $m_N$ (``canonical'' here means that all lepton mixing elements $|U_{\ell N}|$ are factorized out of the expressions). In Fig.~\ref{fig:width:tot} (right) we show the ``canonical'' branching ratios for the full processes $W^+\to N \ell^+ \to n\pi +\ell^+ \ell^+ $ and $W^+\to N\ell^+ \to \bar u d \,\ell^+ \ell^+ $. The actual rates (left) and branching ratios (right) can be obtained by multiplying these canonical values by
$|U_{\ell N}|^2$ (left) and $|U_{\ell N}|^4/\sum_{l^\prime} |U_{\ell^\prime N}|^2$ (right), respectively.
From these figures we were able to estimate the expected number of $W\to N \ell \to n\pi +\ell \ell $ events at the LHC,
%
or equivalently the minimal value of the lepton mixing element   that would generate 5 events or more, for a benchmark value of
$m_N= 10$~GeV.
At this mass, the figure gives a canonical branching ratio
\begin{align}
& {\cal B}(W^+\to \pi^- \mu^+ \mu^+, \pi^0\pi^- \mu^+ \mu^+, \pi^+\pi^-\pi^-\mu^+ \mu^+,\pi^0\pi^0\pi^-\mu^+ \mu^+)\nonumber\\
& \hspace{36pt} \equiv {\cal B}(W^+\to n\pi +\mu^+\mu^+)
 \approx 8\times 10^{-4}.
\end{align}
According to
Ref.~\cite{Aad:2016naf}, at the end of the LHC Run II one may expect a sample of ${\cal N}_W \sim 10^9$ $W$ decays.
Therefore, in order to obtain more than 5 events, we must have:
\[
{\cal N}_W\times {\cal B}(W^+\to n \pi  +\mu^+ \mu^+)  \times \frac{|U_{\mu N}|^4}{\sum_{\ell} |U_{\ell N}|^2 }  > 5 ,
\]
which implies $|U_{\mu N}|^2 \gtrsim 6.2\times 10^{-6}$, provided other mixing elements are smaller. If instead all mixing elements are comparable, then this lower bound increases by a factor 3, i.e.  \hbox{$|U_{e N}|^2,
\, |U_{\mu N}|^2, \, |U_{\tau N}|^2 \gtrsim 1.9 \times 10^{-5}$}. These results are in ideal conditions,  with no cuts or backgrounds.
These bounds can be made about one order of magnitude stronger if one adds both charges and all lepton flavors $W^\pm\to n\pi +\ell^\pm \ell^{\prime \pm}$ $(n=1,2,3)$, i.e.
$|U_{e N}|^2,  \, |U_{\mu N}|^2, \, |U_{\tau N}|^2 \gtrsim 2 \times 10^{-6}$.

One last important point in the observability of these processes is the long lifetime of $N$, which would cause an observable displacement in the detector between the production and decay vertices of $N$. This displacement will drastically help reduce the possible backgrounds. For a sterile neutrino $N$ with mass in the range 5 GeV to 20 GeV, the total width can be estimated as~\cite{Dib:2015oka, Drewes:2019fou}:
\begin{equation}
\Gamma_N \sim \frac{G_F^2 m_N^5}{12 \pi^3} \sum_\ell |U_{\ell N}|^2.
\label{nwidth}
\end{equation}
This expression disregards hadronization effects in the final state; to include hadronization, more detailed calculations~\cite{Bondarenko:2018ptm} change this result by about 12\% in our mass range of interest.
The lighter $N$  and the smaller the mixing $|U_{\ell N}|^2$ are, the longer $N$ will live. Using the current upper bound $|U_{\ell N}|^2\sim 10^{-5}$, the characteristic displacement $\tau c  \equiv \hbar c/\Gamma_N$  is in the range $\sim 20 \ \mu m$ to 20 $mm$ (for 20 GeV and 5 GeV respectively). For smaller $|U_{\ell N}|^2$, the displacements will be proportionally larger. Moreover, the
relativistic $\gamma$ factor will increase the displacement as well.

\section{Detector simulations and Discussion}\label{sec:numerical}

Now we run simulations in order to study the observability of these modes at the LHC. In principle they could be observed provided the pions can be identified and the  background can be reduced using the spatial displacement between the production and decay vertices of the heavy neutrino $N$. This vertex displacement should be observable for $m_N$ below
20 GeV~\cite{Helo:2013esa,Dib:2014iga}.

\begin{figure}[htbp]
   \centering
   \includegraphics[width=0.65\textwidth]{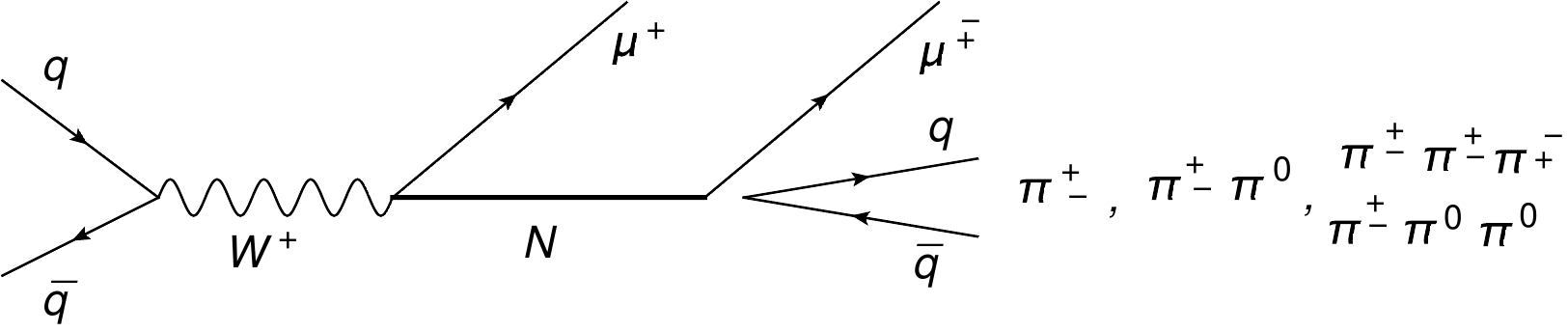} 
   \caption{Heavy-neutrino produced via charged current Drell-Yann process: $pp \rightarrow  W^+ \rightarrow  \mu^+ N (\rightarrow \mu^\mp  + n\ \pi)$. The second muon as $\mu^+$ corresponds to a Majorana $N$; for $\mu^-$ it corresponds to either a Dirac or Majorana $N$.}
   \label{fig:exp1}
\end{figure}

We simulated the whole process shown in Fig.~\ref{fig:exp1}, namely \hbox{$pp\to W\to \mu N, N\to \mu +n\pi$}
at 13 TeV center-of mass energy,
 generating a sample of 15 thousand $W\to \mu N$ events.
We use \textsc{MadGraph5\_aMC@NLO}~\cite{Alwall:2014hca} to generate heavy neutrinos via the charge current Drell-Yan process shown in Fig.~\ref{fig:exp1}. Then, decay and hadronization processes are done with \textsc{Pythia 8.1}~\cite{Sjostrand:2007gs, Sjostrand:2006za}. A fast detector simulation is performed by \textsc{Delphes 3}~\cite{deFavereau:2013fsa}; we use the card ATLAS.tcl included in the package. The UFO~\cite{Degrande:2011ua} files were implemented with Wolfram Mathematica~\cite{Mathematica} using the FeynRules libraries~\cite{Alloul:2013bka}.

As trigger we use events with two muons: one with $p_T > 10$ GeV and the other with $p_T> 4 $ GeV.
Concerning the $W$ and $N$ reconstruction, the events are selected based on the ATLAS/CMS \cite{Aad:2008zzm,Chatrchyan:2008aa} standard requirements: muon $p_{T} > 10$ GeV, $|\eta|<2.5$, and tracks $p_{T} > 500$ MeV and $|\eta| < 2.5$. As the heavy-neutrino mass is considered to be smaller than 20 [GeV], the quarks produced in the neutrino decay are likely to hadronize in the specific pion states described above. We also require events with at least two muons where one of them must show a transverse momentum significantly larger than the other
(see Table I).

\begin{table}[htp]
\caption{Cuts used for the event selection and reconstruction}
\begin{center}
\begin{tabular}{ c || c| c| c }
\hline
&\hspace{6pt} prompt $\mu$ \hspace{6pt}& \hspace{6pt}displaced $\mu$ \hspace{6pt}& \hspace{6pt}displaced tracks (pions) \\
 \hline
$p_T$\hspace{6pt} & $>10$ GeV & $ > 4$ GeV & $>0.5$ GeV
\\
$|\eta|$\hspace{6pt}  & $< 2.5$ & $< 2.5$ & $< 2.5$
\\
\hline
\end{tabular}
\end{center}
\label{default}
\end{table}%

The two muons plus the pions coming from the $N$ decay should reconstruct the $W$ mass. In addition, the less energetic muon plus the pions should
reconstruct the $N$ mass.   However, we must take into account that the neutral pions in the final state will not be detected: they decay immediately into two almost collinear photons which  will not have enough energy to be distinguished from noise in the EM calorimeter in the cases of interest $m_N < 20 $ GeV. Therefore, we can have only two types of events:
($i$) the muons and one charged pion and ($ii$) the muons and three charged pions.
The latter corresponds clearly to $N\to \mu^\pm \pi^\pm \pi^\mp \pi^\mp$, but the former will be the sum of the three decay modes, namely $N\to \mu^\pm \pi^\mp$, $N\to \mu^\pm \pi^\mp \pi^0$ and $N\to \mu^\pm \pi^\mp\pi^0 \pi^0$.

\begin{figure}[htbp]
   \centering
   \includegraphics[width=0.49\textwidth]{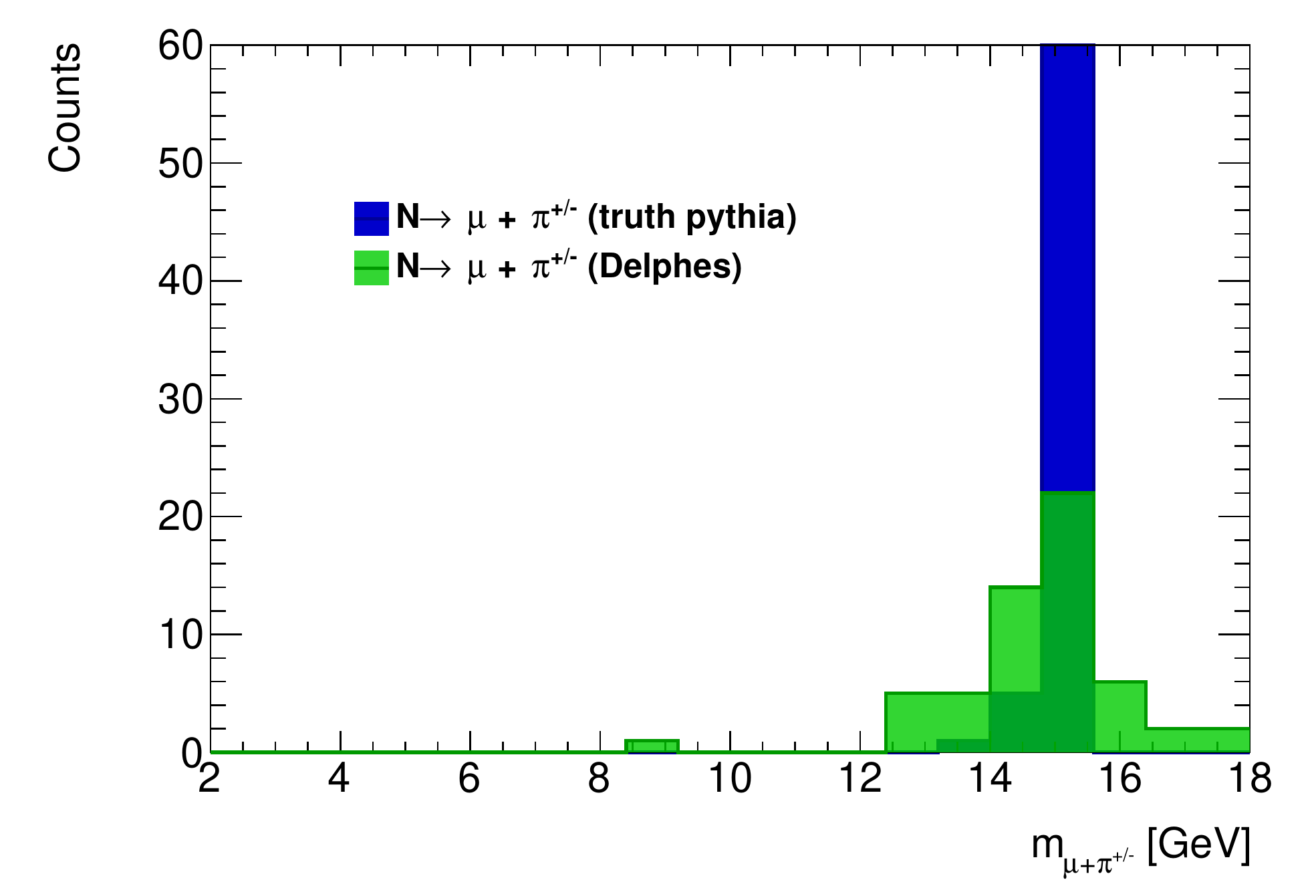}
    \includegraphics[width=0.49\textwidth]{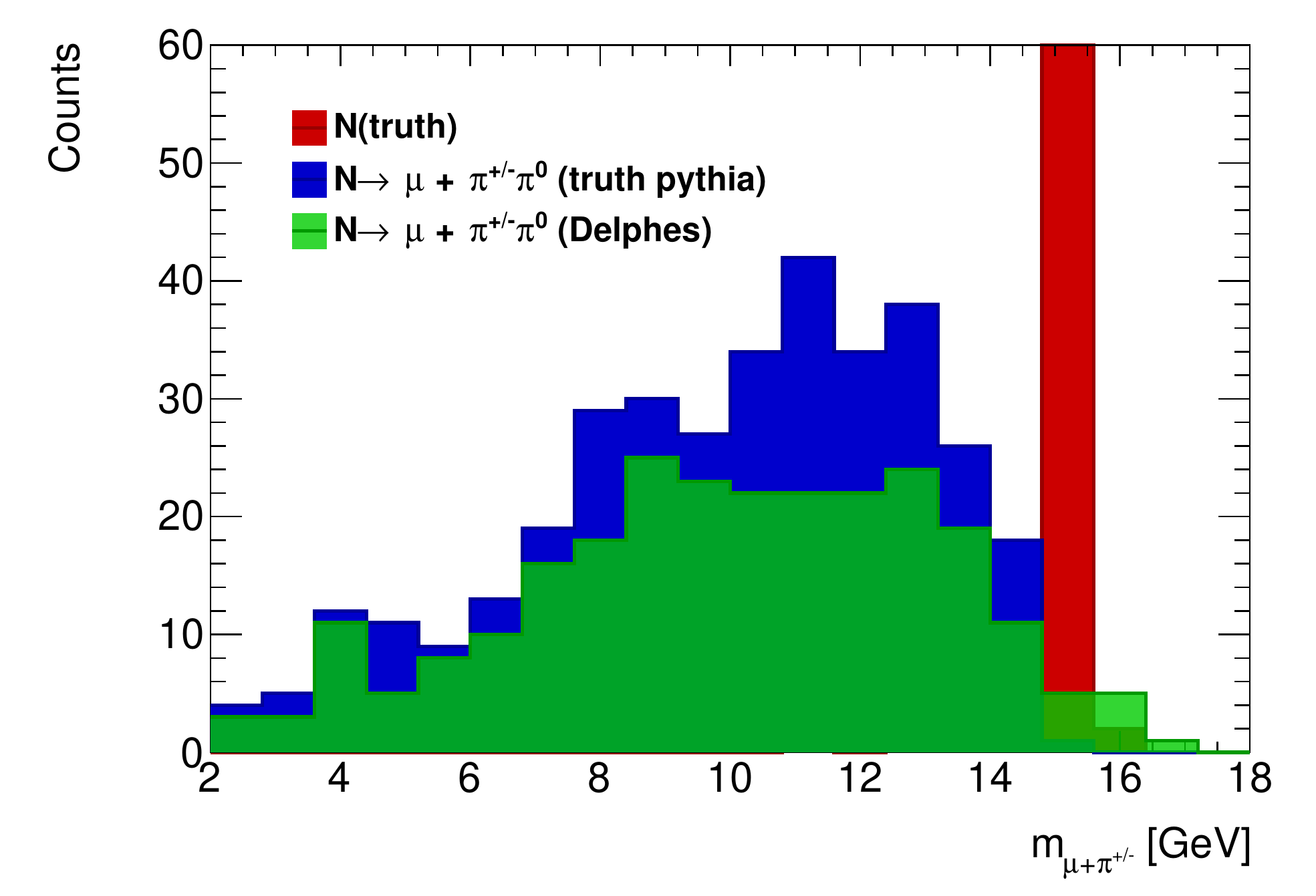}
   \caption{Simulation of a heavy-neutrino reconstructed in hypothetically isolated processes
   $N\to \mu^\pm\pi^\mp$ (left) and $N\to \mu^\pm\pi^\mp\pi^0$ (right) where the $\pi^0$ cannot be detected.  In both cases the invariant mass is of the pair muon-charged pion. Red: idealized case where all particles including neutral pions are detected. Blue: simulation where neutral pions are not detected. Green: same as Blue but smeared by detector resolution (Delphes).
A sample of 15 k events $W\to \mu N$ was used, with a benchmark $m_N= 15$ GeV.}
   \label{fig3}
\end{figure}

In Fig.~\ref{fig3} (left) we show the simulation of the hypothetical case in which the decay into a single pion, $N\to \mu^\pm\pi^\mp$, could be separated from the decays $N\to \mu^\pm \pi^\mp \pi^0,  \mu^\pm \pi^\mp \pi^0 \pi^0$ (this separation is not realistic because neutral pions are not detected). The peak at $m_N$ is clear.  In Fig.~\ref{fig3} (right), we simulate the decay into two pions, where one of the pions has to be neutral due to charge conservation:
$N\to \mu^\pm\pi^\mp\pi^0$. Now, since the $\pi^0$ is not observed, the distribution shows that effect as an extended continuum into lower invariant masses, with an upper cutoff at $m_N$.
\begin{figure}[htbp]
   \centering
   \includegraphics[width=0.49\textwidth]{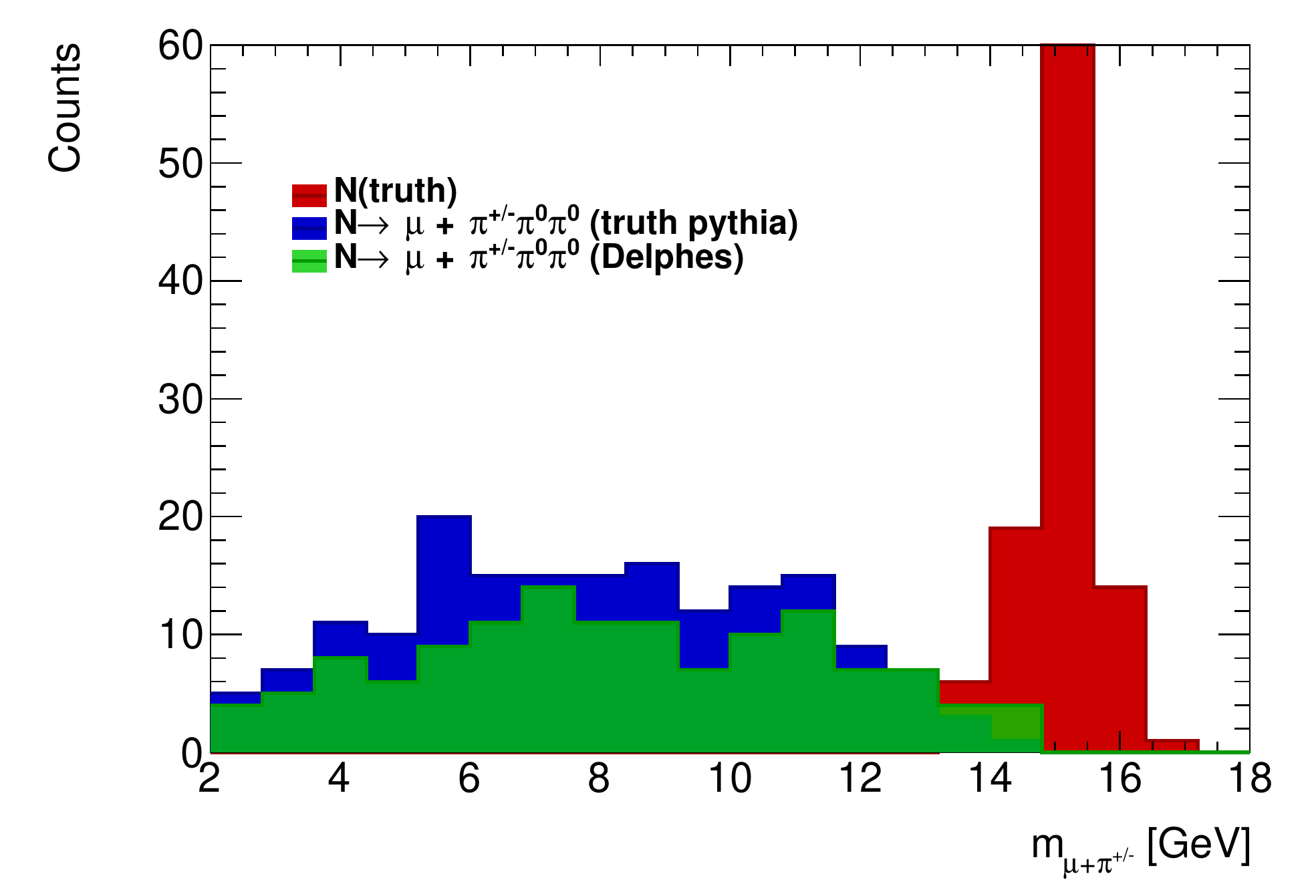}
    \includegraphics[width=0.49\textwidth]{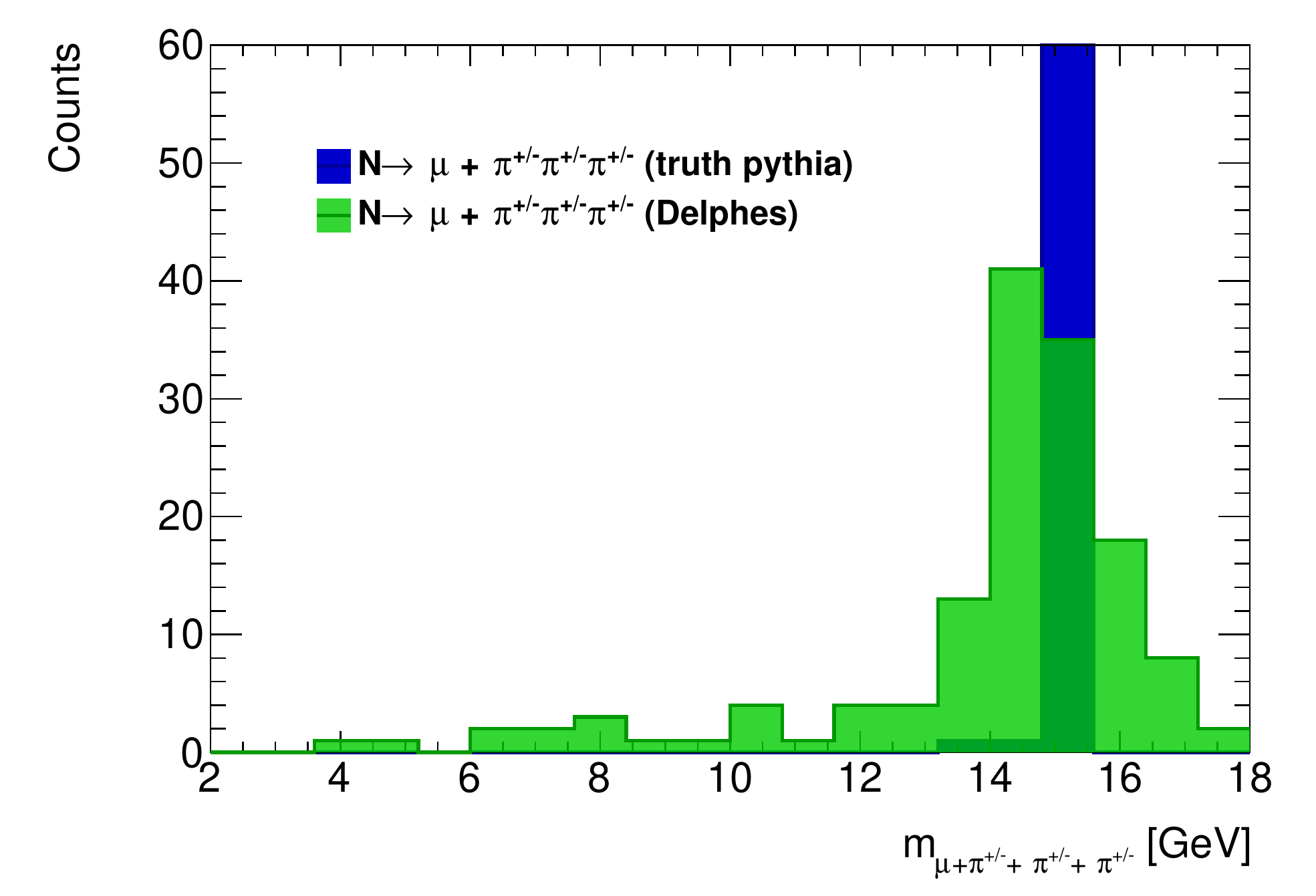}
   \caption{Simulation of a heavy-neutrino reconstructed in hypothetically isolated process
   $N\to \mu^\pm\pi^\mp\pi^0\pi^0$ (left) and in the three-charged pion mode
   $N\to\mu^\pm \pi^\pm\pi^\mp\pi^\mp$.
   Red: idealized case where all particles including neutral pions are detected. Blue: simulation where neutral pions are not detected. Green: same as Red but smeared by detector resolution. A sample of 15 k events $W\to \mu N$ was used, with a benchmark $m_N= 15$ GeV.}
   \label{fig4}
\end{figure}

In Fig.~\ref{fig4} we show the simulations for the decay into three pions.
Now there are two modes: $\mu^\pm\pi^\mp\pi^0\pi^0$ (left) and $\mu^\pm\pi^\pm\pi^\mp\pi^\mp$ (right). In Fig.~\ref{fig4} (left) one can see again a continuous distribution due to the missing $\pi^0\pi^0$, which is even more pronounced to low invariant masses of the charged pair, compared to the single missing $\pi^0$
of Fig.~\ref{fig3} (left).  In contrast, the three-charged pion mode shows in Fig.~\ref{fig4} (right)
shows a clear reconstruction of the peak at $m_N$.

\begin{figure}[ht]
   \centering
   \includegraphics[width=0.49\textwidth]{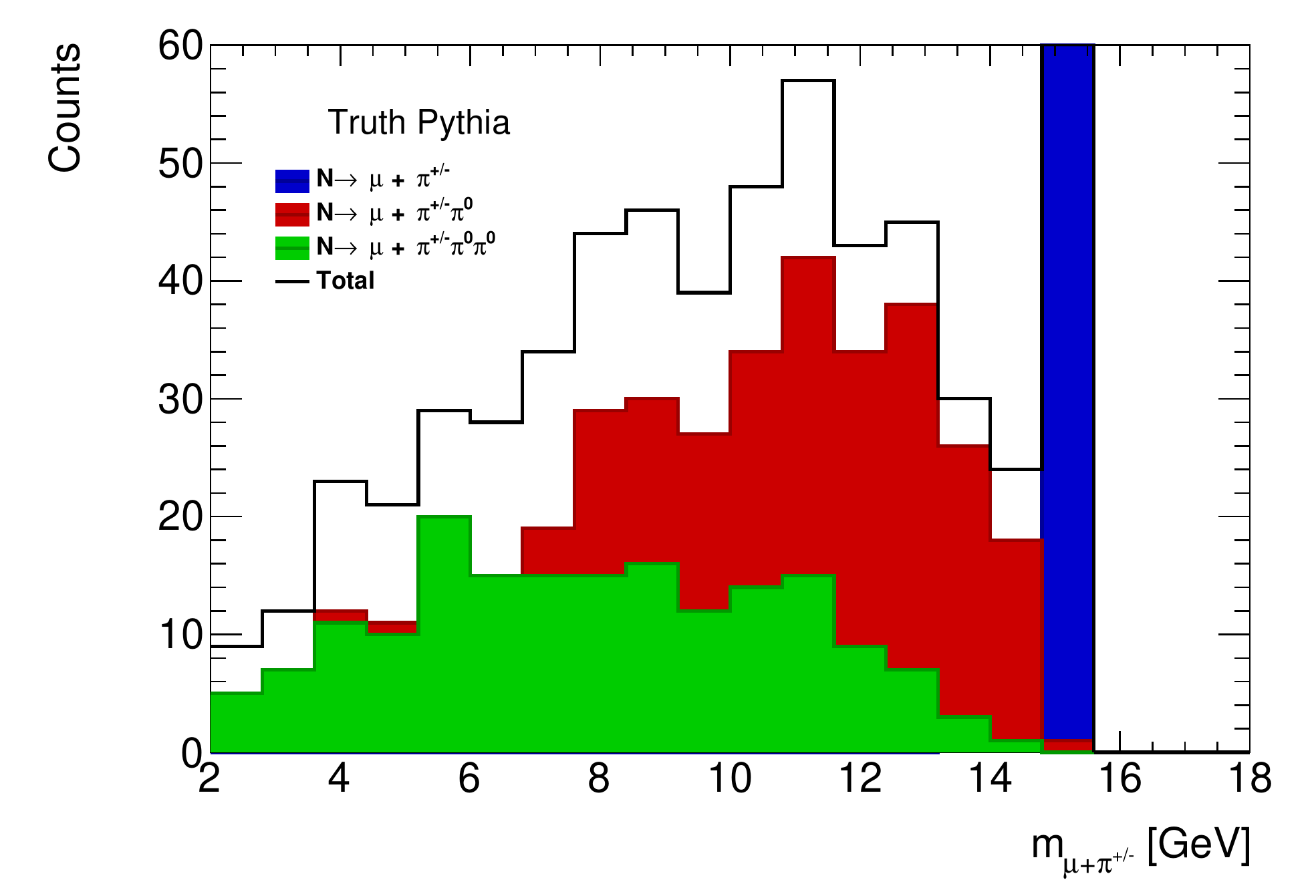}
    \includegraphics[width=0.49\textwidth]{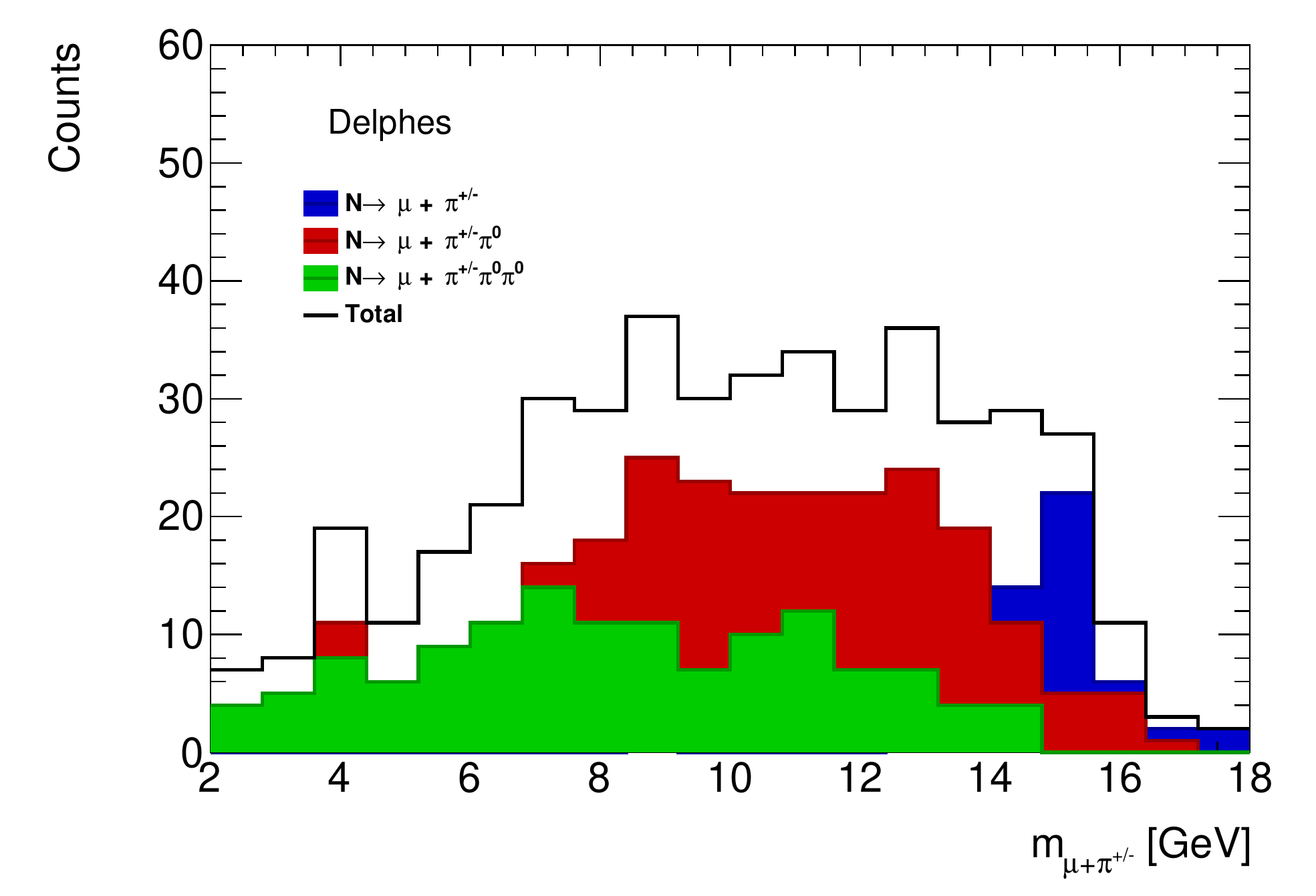}
   \caption{Simulation of a heavy neutrino reconstructed in events with a muon and a single
   charged pion. The events combine the decays into $\mu^\pm\pi^\mp$,
   $\mu^\pm\pi^\mp\pi^0$ and $\mu^\pm\pi^\mp\pi^0\pi^0$ where the neutral pions are missing.
   A sample of 15 k events $W\to \mu N$ was used, with a benchmark $m_N= 15$ GeV.}
   \label{fig5}
\end{figure}

In a realistic case, therefore, the search for a mode with a single charged pion will be the sum of the three pionic final states, $\pi^\pm$, $\pi^\pm\pi^0$ and $\pi^\pm \pi^0\pi^0$, as we show in the simulations of Fig.~\ref{fig5} (there are modes with even more pions, but those are suppressed with respect to the three cases considered here). Fig.~\ref{fig5} (left) shows the result from \textsc{Pythia} and Fig.~\ref{fig5} (right) the smeared distribution due to detector resolution given by \textsc{Delphes}. The latter is a more spread distribution, but qualitatively they are similar in the sense that the dominant feature is a continuous distribution with an upper endpoint at the $N$ mass. The single pion decay channel with no neutral pions, namely $N\to \mu^\pm\pi^\mp$, which is the only mode with a clear peak, is suppressed compared to the other two channels and its peaked feature is lost in the distribution. This feature contrasts with the
events with 3 charged pions, shown in Fig.~\ref{fig4} (right), where the peak is still clear.
In a real search there should be a continuous part due to the 4-pion mode, but it is subdominant.

\begin{figure}[ht]
   \centering
   \includegraphics[width=0.49\textwidth]{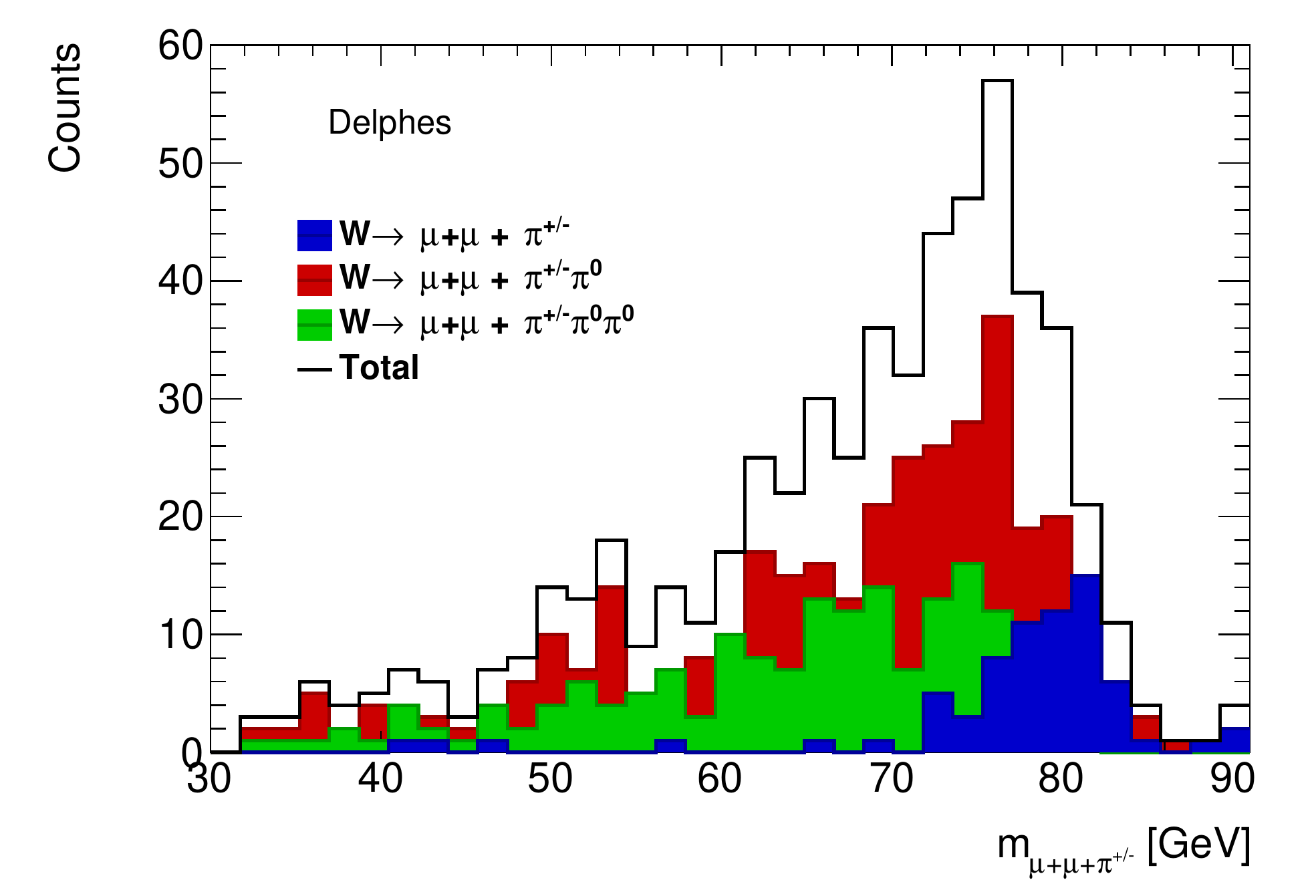}
   \includegraphics[width=0.49\textwidth]{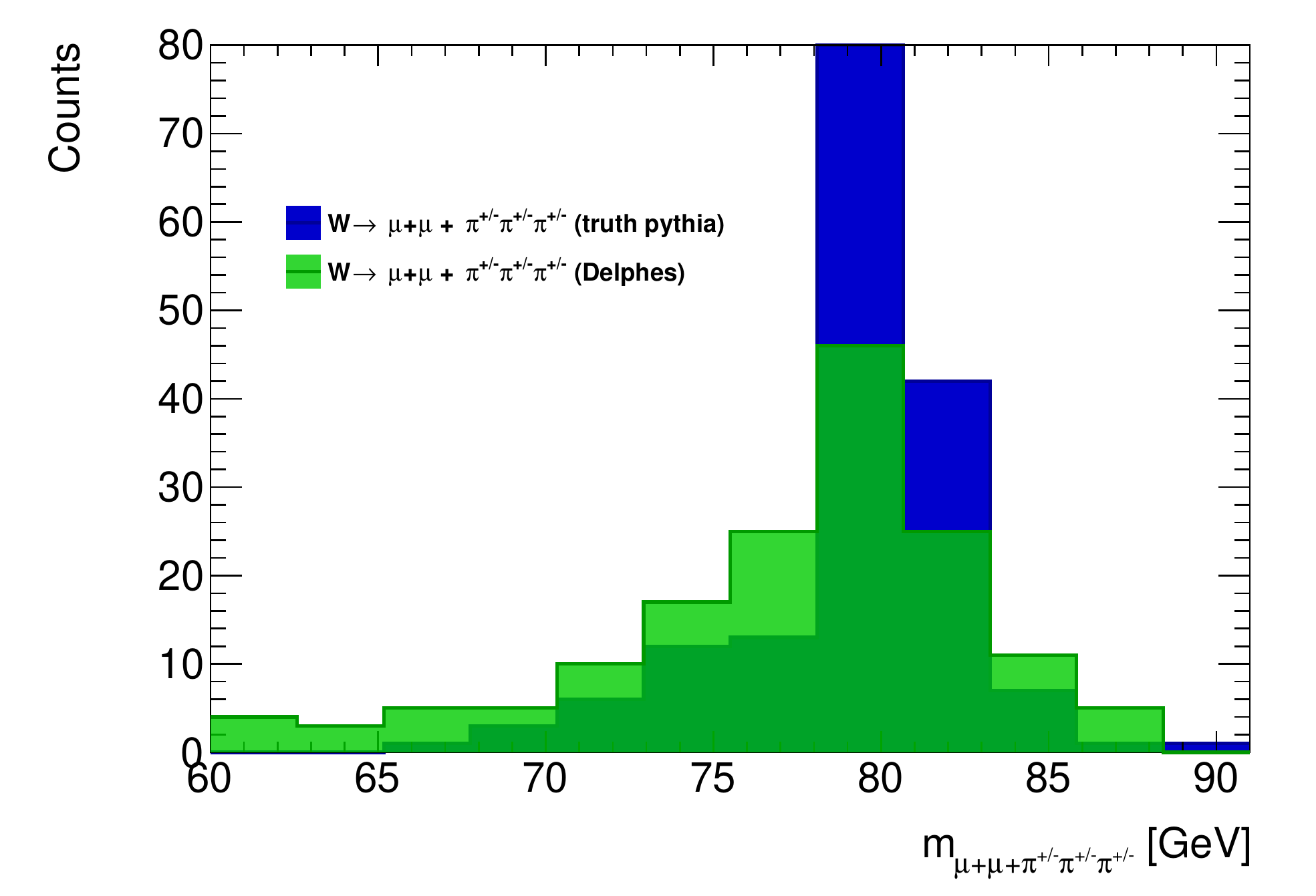}
   \caption{Simulation of the $W$ boson decaying into $\mu N$ followed by the $N$ decay, reconstructed in events with two muons and a single
   charged pion (left), and three charged pions (right).
   A sample of 15 k events $W\to \mu N$ was used, with a benchmark $m_N= 15$ GeV.}
   \label{fig6}
\end{figure}

Besides the reconstruction of the $N$, one should verify the reconstruction of the $W$
from the full event, which essentially adds the prompt muon to the decay of $N$. Given that we are considering $m_N< 20$ GeV, i.e. considerably lighter than $M_W$, the prompt muon should be more energetic that the one coming for the $N$ decay and, due to the relatively long lifetime of an $N$ with such masses,  the decay vertex of $N$ should be displaced with respect to the
prompt muon.
Figs.~\ref{fig6} show the distributions for the $W$ reconstruction in the respective modes for the $N$ decays described above, namely with a single charged pion (Fig.~\ref{fig6} left) and in the mode with three charged pions (Fig.~\ref{fig6} right).

Concerning the reconstruction efficiencies at the LHC for our particles of interest, these constitute no major limitation in the results, since for muons it is above 98\% \cite{Aad:2016jkr},  while for pions the tracking efficiency is near 90\% \cite{Aaboud:2017all} and
\cite{ATL-PHYS-PUB-2015-051}.
For the displaced vertex efficiency one should expect near 50\% for 4 charged tracks \cite{ATL-PHYS-PUB-2019-013}, although we did not 
require vertex displacement in the analysis to build Figs. 3-6.
As stated at the end of Section II, the characteristic displacement due to the long $N$ lifetime could be from a few tens of micrometers to a few meters, depending on $m_N$ and the mixing. This implies that a fraction of the decays occurs outside the range of observability of the detector.  If we consider a minimum and maximum observable displacement, $L_0$ and $L_1$ respectively, the acceptance factor (the probability that the decay occurs within the fiducial volume of the detector) will be
\begin{equation}
ac= (e^{-\gamma L_0/\tau c} - e^{-\gamma L_1/\tau c}).
\label{accept}
\end{equation}
The efficiency and acceptance factors should be determined by the specific experiment using the data and detector, so we did not include them in the previous analysis.

\begin{figure}[ht]
   \centering
   \includegraphics[width=0.49\textwidth]{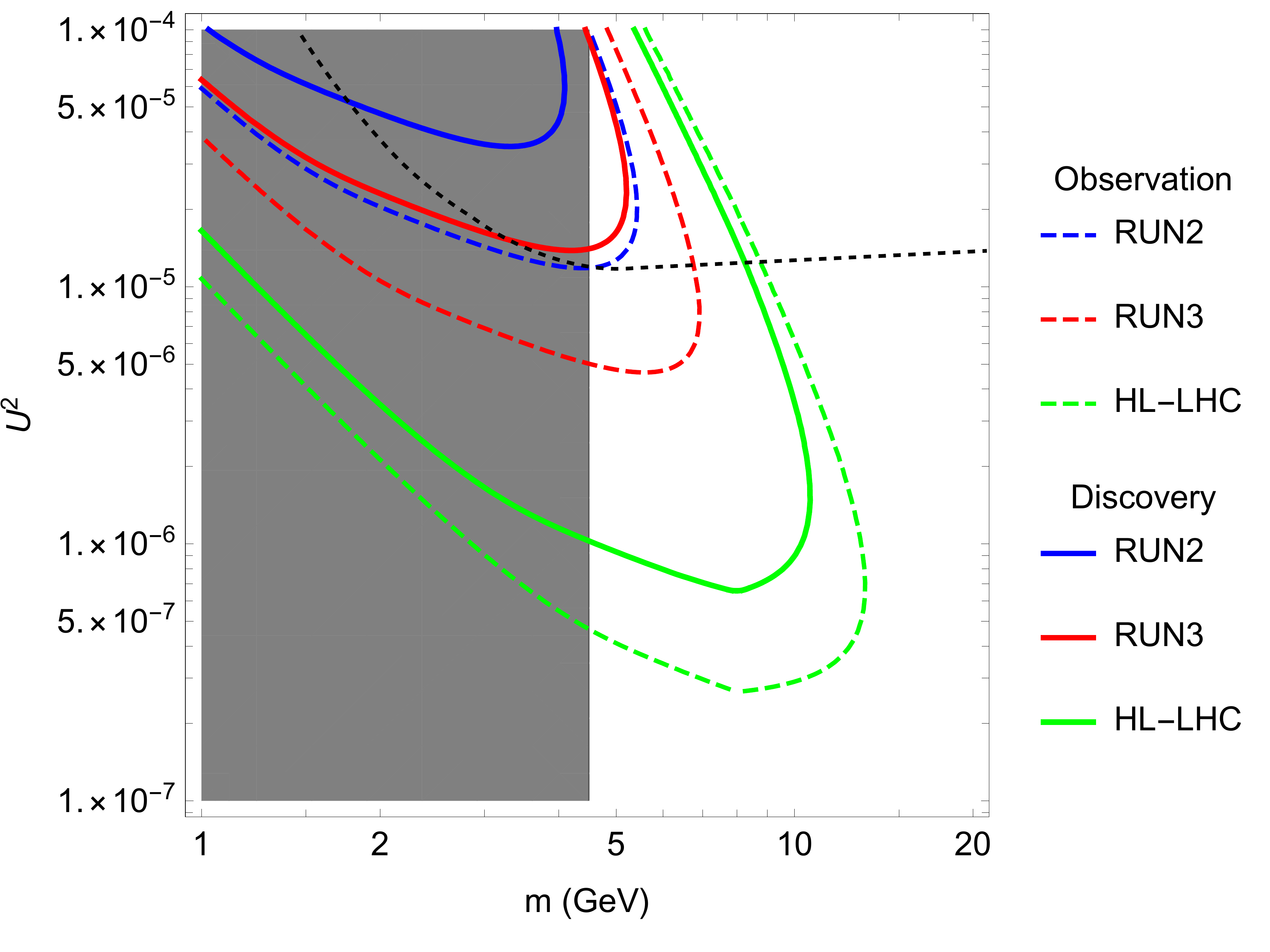}
   \includegraphics[width=0.49\textwidth]{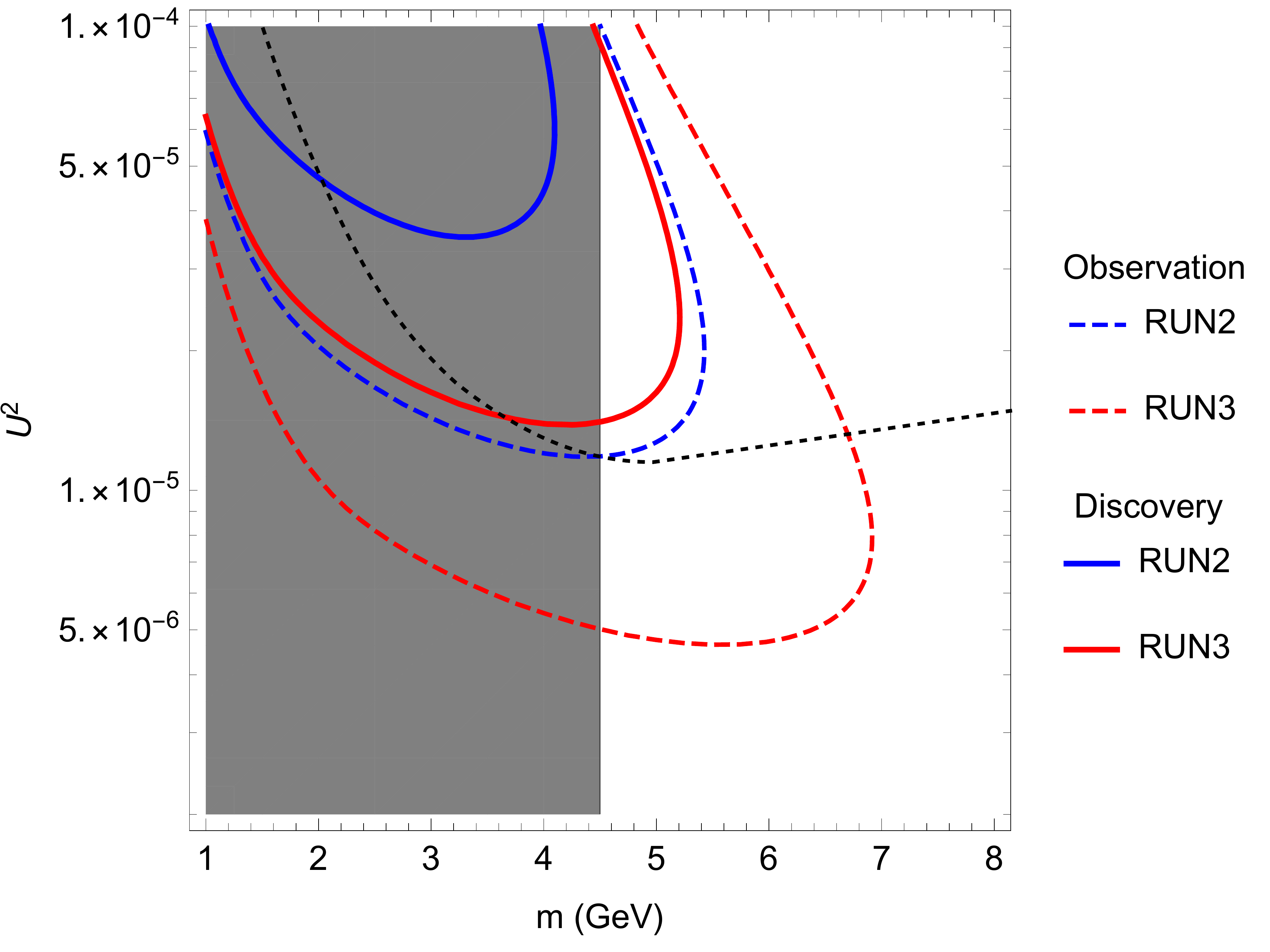}
   \caption{Sensitivity of $pp\to \mu N (N\to n\pi \mu)$ to the mixing $|U_{\mu N}|^2$, as a function of the sterile neutrino mass, $m_N$, for the LHC Run 2 (150 fb$^{-1}$), Run 3 (300 fb$^{-1}$) and High-Luminosity LHC (3000 fb$^{-1}$).
A vertex displacement detectability between 1 mm and 30 cm is assumed.
 The dashed curves correspond to 4 events for observation and the solid curves to 9 events for discovery.
 The black dotted line is the current bound from DELPHI \cite{Abreu:1996pa}. The grey region is affected by backgrounds not considered in the analysis and so the limits are less reliable there.
 }  \label{sensitive}
\end{figure}

Nevertheless, in order to have an estimate of the potential sensitivity of these pionic modes to the neutrino mixing $|U_{\mu N}|^2$, we have simulated the events for assumed values $L_0 = 1$ mm and $L_1 = 30 $ cm.
The sensitivity curves  for the LHC Run 2, the LHC Run 3 and the LH-LHC are shown in Fig.~\ref{sensitive}.  The curves show the smallest values of $|U_{\mu N}|^2$ to observe 4 events (dashed lines) or 9 events (solid lines), which would correspond to $3\sigma$ and $5\sigma$, respectively, assuming 1 background event \cite{Cowan:2010js}. The current limit by DELPHI \cite{Abreu:1996pa} is also shown.

The sensitivity curves also include
the cuts in Table 1,
a 50\% efficiency factor for muon, tracking and DV (displaced vertex) detection, and the DV acceptance factor of Eq.~\eqref{accept} with the assumed values of $L_0$ and $L_1$ associated to the detector,
and the average value of the relativistic factor $\gamma$ for each given mass.
The curves would extend to lower $m_N$ values if the detector length $L_1$ is larger than the assumed value of 30 cm, and would extend to larger $m_N$ values if the smallest observable displacement $L_0$ is smaller than the assumed value of 1 mm.

From Fig.~7 one can see that Run 2 does not seem to improve on the current bound imposed by DELPHI, while Run 3 has a potential to tighten the bound on the mixing by an order of magnitude for $m_N$ in the low mass range (around 5 to 7 GeV). In contrast, the High-Luminosity LHC (HL-LHC) could improve the bounds by about 2 orders of magnitude in the mass range below 11 GeV.

The figures above are obtained from simulations of the signal only.
 In order to eliminate backgrounds in the actual experiment, one should also request:
(a)
Minimal amount of missing $E_T$: with this cut, one would remove all events that have SM neutrinos, such as $W^+W^-$ or $t \bar t$ production.
%
(b)
Displaced vertex associated with the muon with lower $p_T$. As stated before, for $m_N < 20$ GeV the prompt muon is more energetic that the second muon, and the latter should
come from a displaced secondary vertex, because the decaying $N$ is rather long living.
This cut should
remove almost all remaining backgrounds with the exception of
heavy-flavors, e.g $B$-hadrons, that typically decay into one displaced muon plus tracks.
%
(c)
Isolation in the prompt muon in addition to the requirement that
the invariant mass of the two muons plus the tracks should be close to the $W$ mass.

In general the background-free hypothesis with displaced vertices can be trusted only for masses larger than about 5 GeV, where no $N$ is produced from meson decays. Consequently, Fig.~7 appears in grey for masses below 4.5 GeV.
For masses below this boundary, one can largely reject backgrounds using the cuts a), b), and c), because all mesons that may lead to displaced vertices are most of the time produced within jet fragments. In particular, cut c), i.e. ``muon isolation'', should remove largely those backgrounds, and remove even more by adding the cut of W invariant mass of the full system.
One may expect that no SM background should remain after cuts a), b) and c),
but to quantify this statement is an issue that goes beyond this work.
Concerning backgrounds due to nuclear interactions with the detector material \cite{Cottin:2018kmq}, these are largely suppressed by the requirement of having one muon to be part of the displaced vertex reconstruction.

\section{Conclusions}\label{sec:conclusion}

In this work we have studied the observability at the LHC of  the exclusive process
\hbox{$W \to \mu N (\to \mu +n\pi)$}, which is appropriate to discover a sterile neutrino $N$ with mass in the range 5 GeV $< m_N < 20$ GeV. This is an intermediate region where neither rare meson decays ($B$, $D$, etc.) nor $pp\to \ell\ell jj, \ell\ell\ell \nu$ modes at the LHC  are sensitive to the presence of such neutrinos. The modes we use are exclusive semileptonic, containing pions in their hadronic component. Because of the pions in the final state, the reconstruction of the events at a hadron collider is not a trivial matter. However, one particular feature of this process that helps reducing drastically all backgrounds is the fact that a sterile neutrino $N$ with mass below 20 GeV should have a lifetime long enough to travel an observable distance in the detector before it decays. Indeed, given the current upper bounds on the sterile neutrino mixing with the muon flavor $|U_{\mu N}|^2\lesssim 10^{-5}$, a vertex displacement above 20 $\mu m$ would occur for $m_N = 20$ GeV, and longer than 20 $mm$ for $m_N = 5$ GeV.

Na\"\i vely, the most favorable mode should have a single charged pion in the final state, namely $N\to\mu^\pm\pi^\mp$. However this mode is suppressed compared to the two-pion and three-pion modes. Moreover, since neutral pions would go undetected at the LHC, the single charged pion events will contain the modes $N\to \mu^\pm\pi^\mp \pi^0$ and $N\to \mu^\pm\pi^\mp \pi^0\pi^0$ as well. We have simulated the events that contain one prompt muon, followed by a displaced muon and charged pion (the second muon should also have less $p_T$ than the prompt muon). The invariant mass of the displaced charged particles $\mu^\pm\pi^\mp$ will show a continuous distribution with an upper endpoint at the $N$ mass, and the invariant mass of all three charged particles $\mu\mu\pi$ will show a continuous distribution with an endpoint at $M_W$.

Accordingly, the cleanest mode is the one with three charged pions in the final state:
\hbox{$N\to \mu^\pm\pi^\pm\pi^\mp\pi^\mp$}. Modes with more pions are suppressed compared to this one, so we have neglected their effect in the invariant mass distributions (one would expect only a small continuous tail to lower invariant masses, due to the modes with additional neutral pions that go undetected). In this mode, the invariant mass of the secondary muon and the charged pions will show a peak at $m_N$ and the invariant mass that includes the prompt muon will show a peak at $M_W$.

These exclusive semileptonic processes with their feature of vertex displacement
can improve the current upper bounds on the mixing of sterile neutrinos with the muon flavor. The LHC Run 3 can improve the bound by an order of magnitude for $m_N$ only in the lower end (5 GeV to  7 GeV), while an improvement about 2  orders of magnitude in the bound can be achieved with the High-Luminosity LHC  for $m_N$ below 11 GeV, where our proposed semileptonic modes are advantageous over other processes. For masses above 11 GeV these semileptonic modes do not seem to give an improvement in the bound, unless the detector is able to distinguish vertex displacements shorter than 1 mm as it was assumed in these simulations.

\section*{Acknowledgements}

C.S.K. was supported in part by the National Research Foundation of Korea
(NRF) grant funded by the Korean government (MSIP) (NRF-2018R1A4A1025334).
S.T.A was supported in part by the National Science Fundation (NSF) grant 1812377.
This work was supported by FONDECYT (Chile) grant 1170171
and CONICYT (Chile) PIA/Basal FB0821.

\bibliography{ntopibib}

\end{document}